\documentclass[
    ,final            
  ]
  {aipproc}

\layoutstyle{6x9}

\def\Journal#1#2#3#4{{#1} {\bf #2}, (#4) #3}
\def\EPJ{Eur.~Phys.~J}

\def\PLB{Phys. Lett. B}
\def\PRL{Phys. Rev. Lett.}
\def\PRD{Phys. Rev. D}

\begin{document}

\title{Quark-hadron duality and higher twist contributions in structure
       functions}

\author{\underline{A.~Fantoni}}{
  address={Laboratori Nazionali dell'INFN, Via E.Fermi 40, 00044 Frascati (RM)
           Italy}
}

\author{N.~Bianchi}{
  address={Laboratori Nazionali dell'INFN, Via E.Fermi 40, 00044 Frascati (RM)
           Italy}
}

\author{S.~Liuti}{
  address={University of Virginia, Charlottesville, Virginia 22901, USA}
}

\begin{abstract}
 The quark-hadron duality is studied in a systematic way for polarized and
unpolarized structure functions, by taking into account all the available data 
in the resonance region.
In both cases, a precise perturbative QCD based analysis to the integrals
of the structure functions in the resonance region has been done: non 
perturbative contributions have been disentangled, the higher twist 
contributions have been evaluated and compared with the ones extracted in the DIS
region. 
A different behavior for the unpolarized and polarized structure
functions at low $Q^2$ has been found.
\end{abstract}

\maketitle

The structure and the interaction of hadrons is generally described by two
different but complementary approaches: the quark-gluon context at 
high energy, where the quarks are asymptotically free, and the meson and 
baryon description at low energy, where the effects of the confinement 
are large.
In some specific cases where the natural description in terms of hadrons 
should be applied,
the quark-gluon description can be successfully also used.   
This evidence is called quark-hadron duality and it was introduced by Bloom 
and Gilman who noted a relationship between the nucleon resonance region and the 
deep-inelastic-scattering (DIS) one~\cite{BG}.

A quantitative analysis of the $Q^2$ dependence of quark-hadron duality
in both polarized and unpolarized $ep$ scattering is presented.
All current data in the resonance region, $1 \leq W^2 \leq 4$ GeV$^2$,
have been taken into account.
For the unpolarized case it has been used the data obtained at Jefferson
Lab in the range $0.3 \leq Q^2 \leq 5$ GeV$^2$ \cite{CEBAF}, and the data
from SLAC (\cite{whit} and references therein) for $Q^2 \geq 4$ GeV$^2$.
For the polarized case there are only few experimental data in the
resonance region.
One set is part of the E143 data \cite{E143}, and it corresponds to
$Q^2 =0.5$ and $1.2$ GeV$^2$.
Another set is the one from HERMES \cite{HERMESPRL,Ale} in the range
$1.2 \leq Q^2 \leq 12$ GeV$^2$.

The full procedure of the analysis is described in \cite{BFL}.
The quark-hadron duality in DIS is studied by considering the ratio of the
integrals of the structure functions integrated in a defined $x$-range,
corresponding to the $W$ range of the resonance region. The structure function
in the numerator is evaluated using the experimental data in the resonance
region, while the one at the denominator is calculated from parametrizations
that reproduce the DIS behavior of the data at large $Q^2$.
The ratios $R_{\mathrm{unpol}}$ and $R_{\mathrm{pol}}$ have been calculated in 
unpolarized and polarized cases, respectively.

In order to understand the nature of the remaining $Q^2$ dependence that cannot
be described by NLO pQCD evolution, the effect of target mass corrections and
large $x$ resummation have been studied.
The analysis was performed by using $x$ as an integration variable, which avoids 
the ambiguities associated to the usage of other {\it ad hoc} kinematical 
variables.
Standard input parametrizations with initial scale $Q_o^2 = 1 $ GeV$^2$ have
been used. Once both effects have been subtracted from the data, and assuming 
the validity of the twist expansion, one can interpret any remaining 
discrepancy of the ratio from unity in terms of higher twist.

The Target Mass Corrections (TMC) are necessary to take into account the 
finite mass of the initial nucleon. 
They are corrections to the leading twist (LT) part of the unpolarized 
structure function $F_2$.

Large $x$ Resummation (L$x$R) effects arise formally from terms containing 
powers of $\ln (1-z)$, $z$ being the longitudinal variable in the evolution 
equations, that are present in the Wilson coefficient functions $C(z)$.
The logarithmic terms in $C_{NS}(z)$ become very large at large $x$, and they 
need to be resummed to all orders in $\alpha_S$~\cite{Rob}. 

\begin{figure}
  \includegraphics[height=.3\textheight]{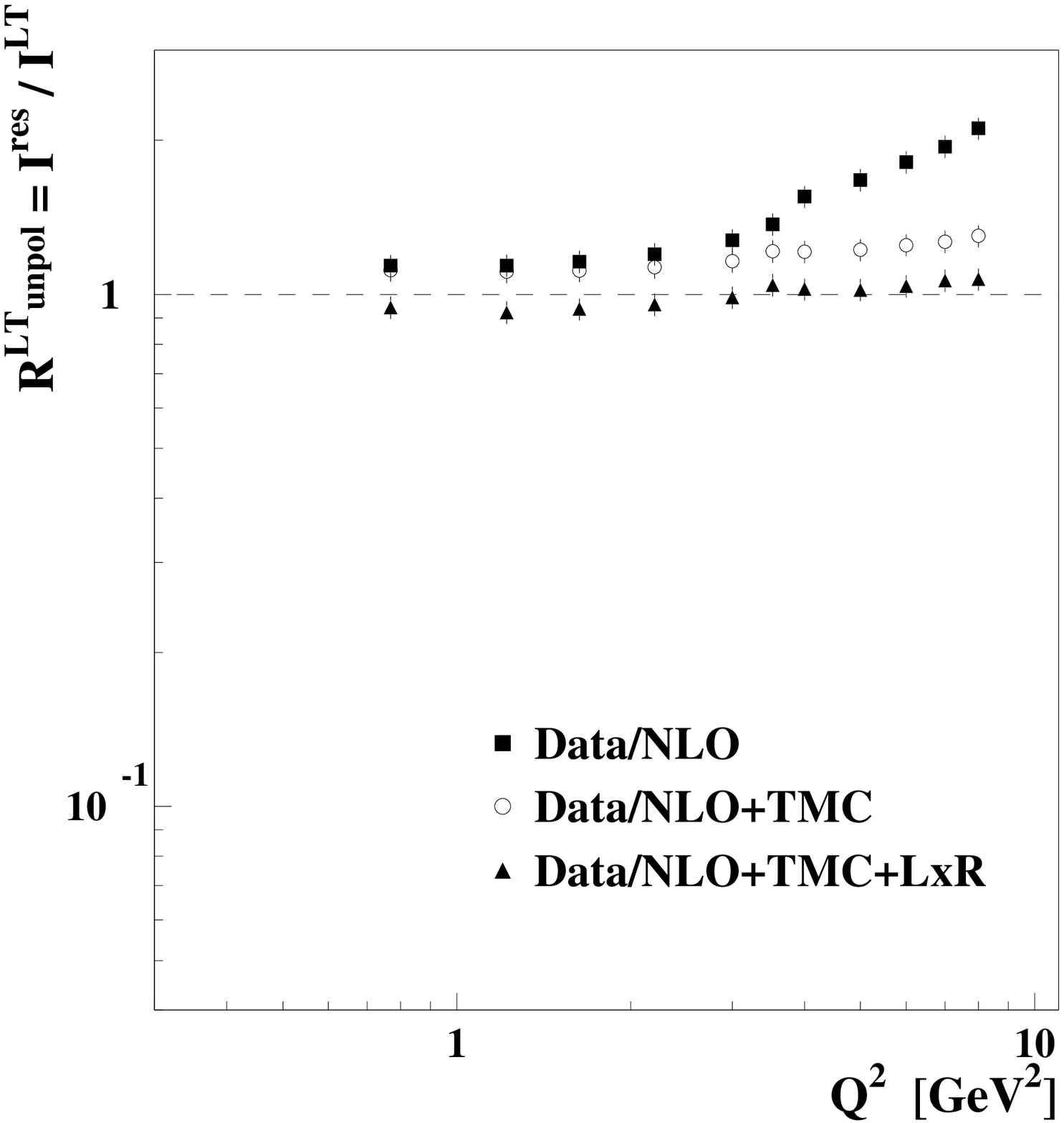}
  \includegraphics[height=.3\textheight]{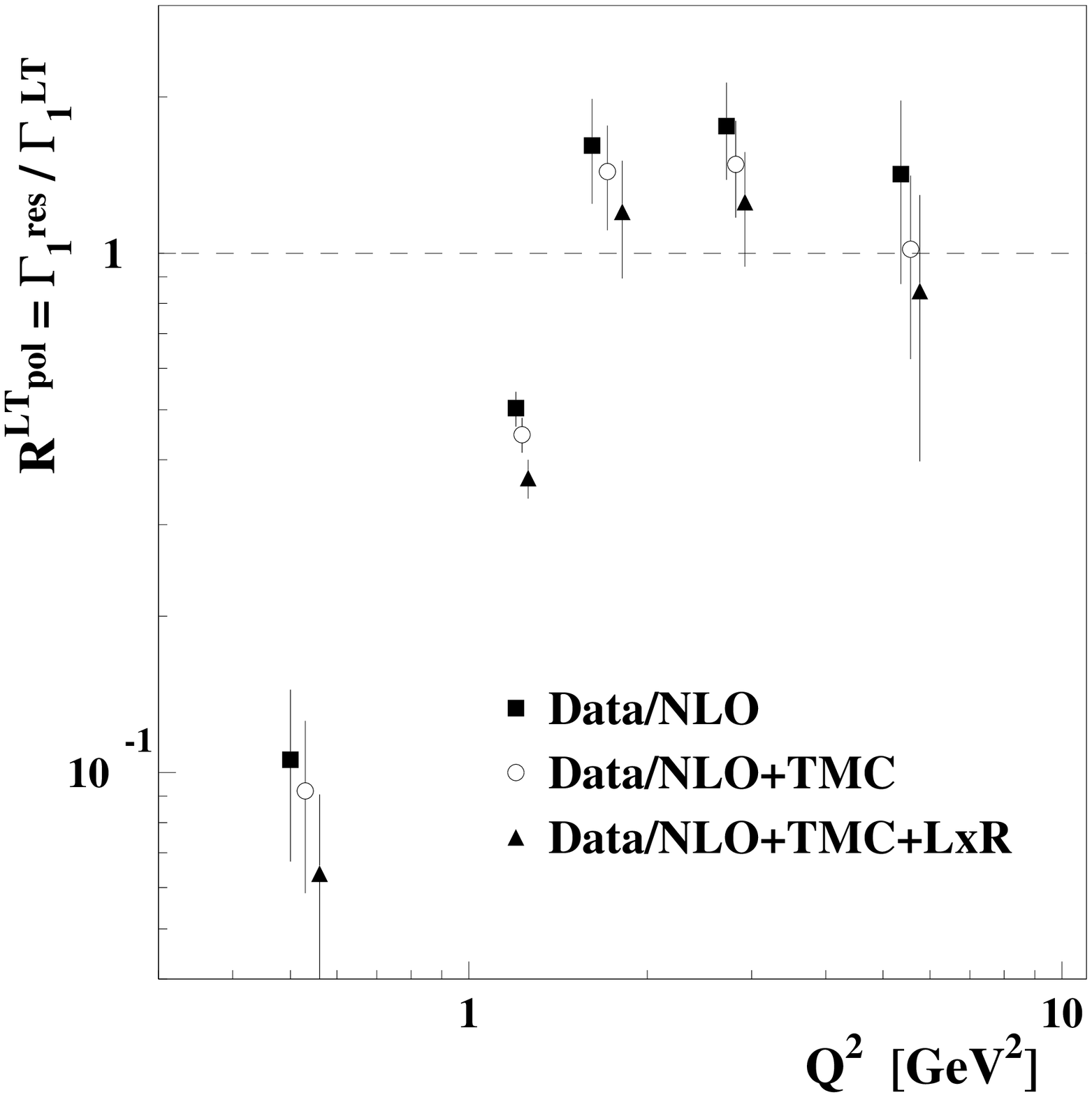}
\caption{Ratio between the integrals of the measured structure functions
         and the calculated ones plotted as a function of $Q^2$ for the
         unpolarized case (left) and the polarized one (right).
         The calculation includes one by one the effects of NLO pQCD.}
\label{theor1}
\end{figure}

All the effects are summarized in Fig.\ref{theor1}, where the ratio between 
the resonance region and the 'DIS' one is reported for the unpolarized and for 
the polarized case: the numerator is obtained from the experimental data, 
while the denominator includes the different components of the present 
analysis, one by one.

\begin{figure}
  \includegraphics[height=.3\textheight]{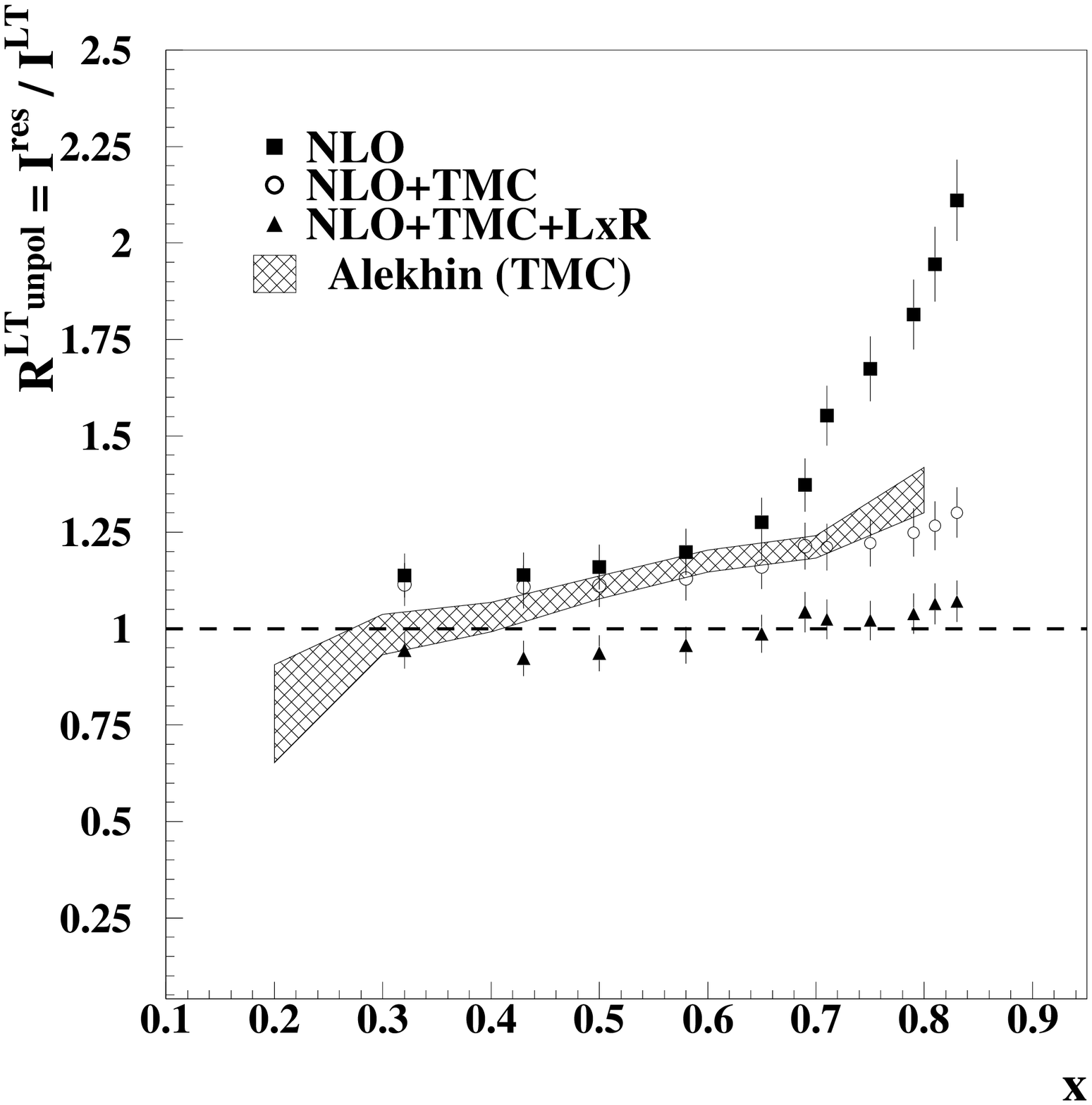}
  \includegraphics[height=.3\textheight]{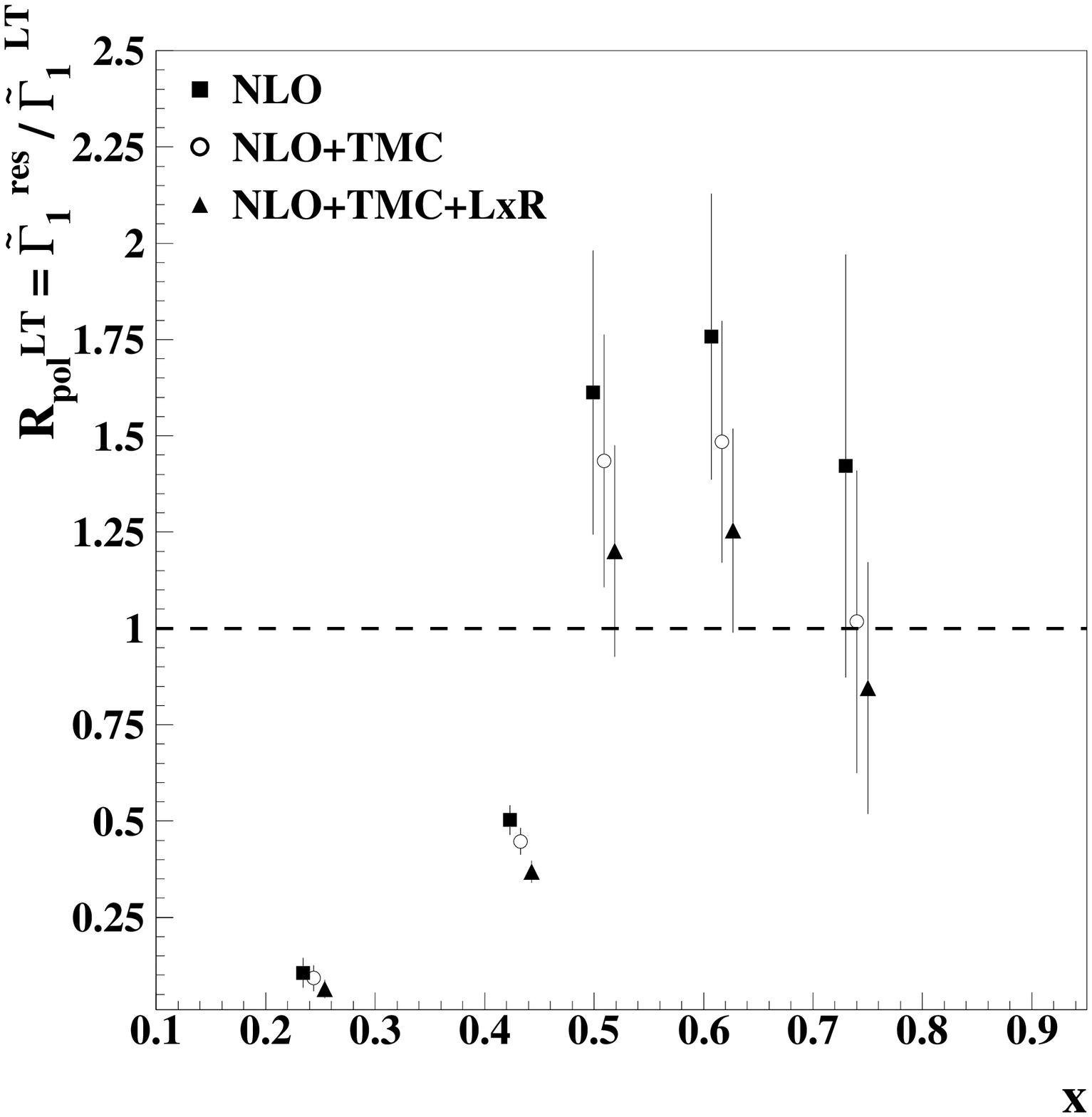}
\caption{HT coefficients extracted in the resonance region according
         to Eq.(\protect\ref{CHT}). Shown in the figure is the quantity
         $1+C_{HT}(x)/Q^2$. For comparison the values for HT coefficients
         obtained in ref.~\cite{Alekhin} using DIS data and the effect
         of TMC are shown.
         The left (right) panel refers to the unpolarized (polarized) case.}
\label{theorx}
\end{figure}

\begin{figure}
  \includegraphics[height=.3\textheight]{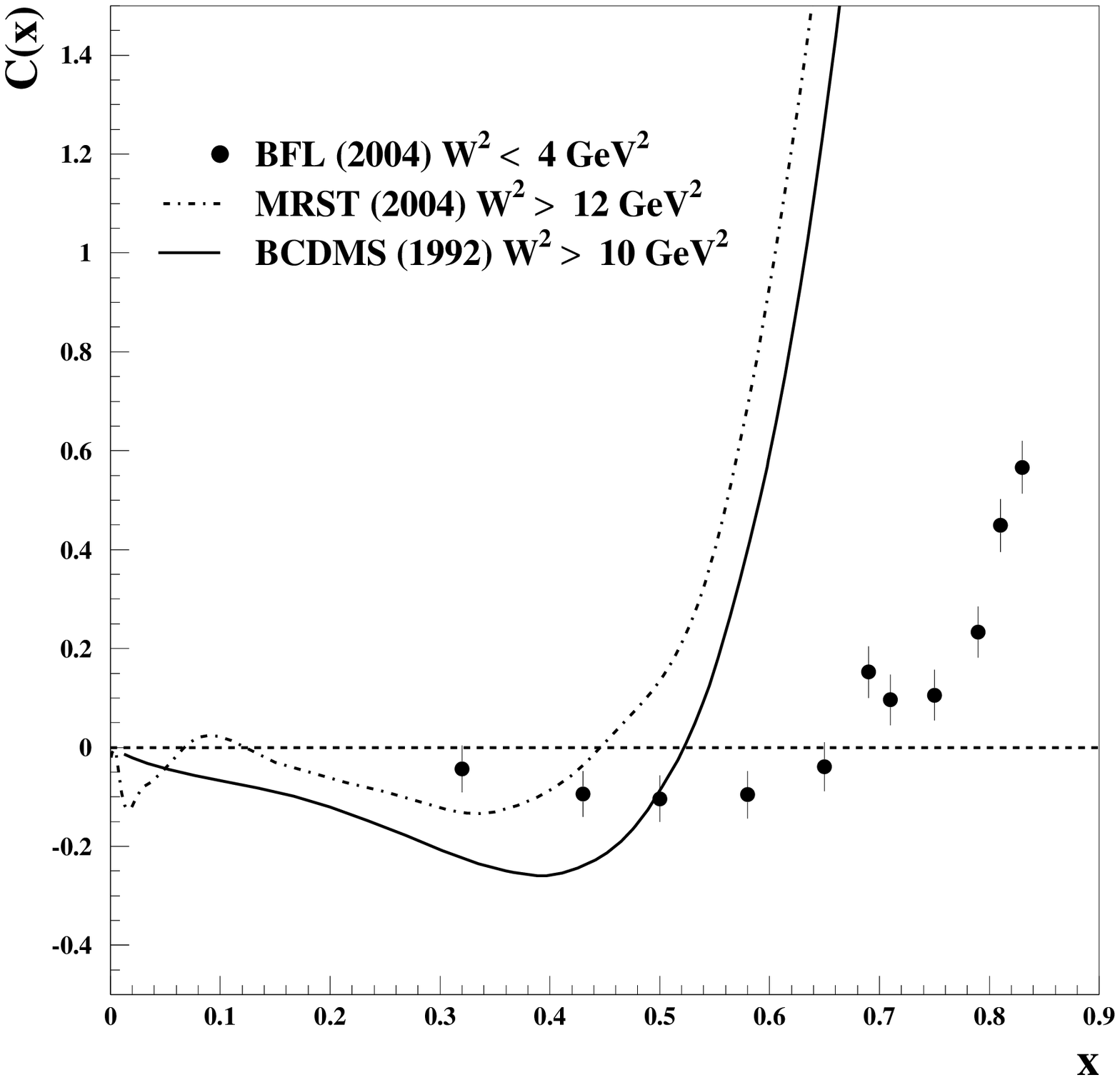}
  \includegraphics[height=.3\textheight]{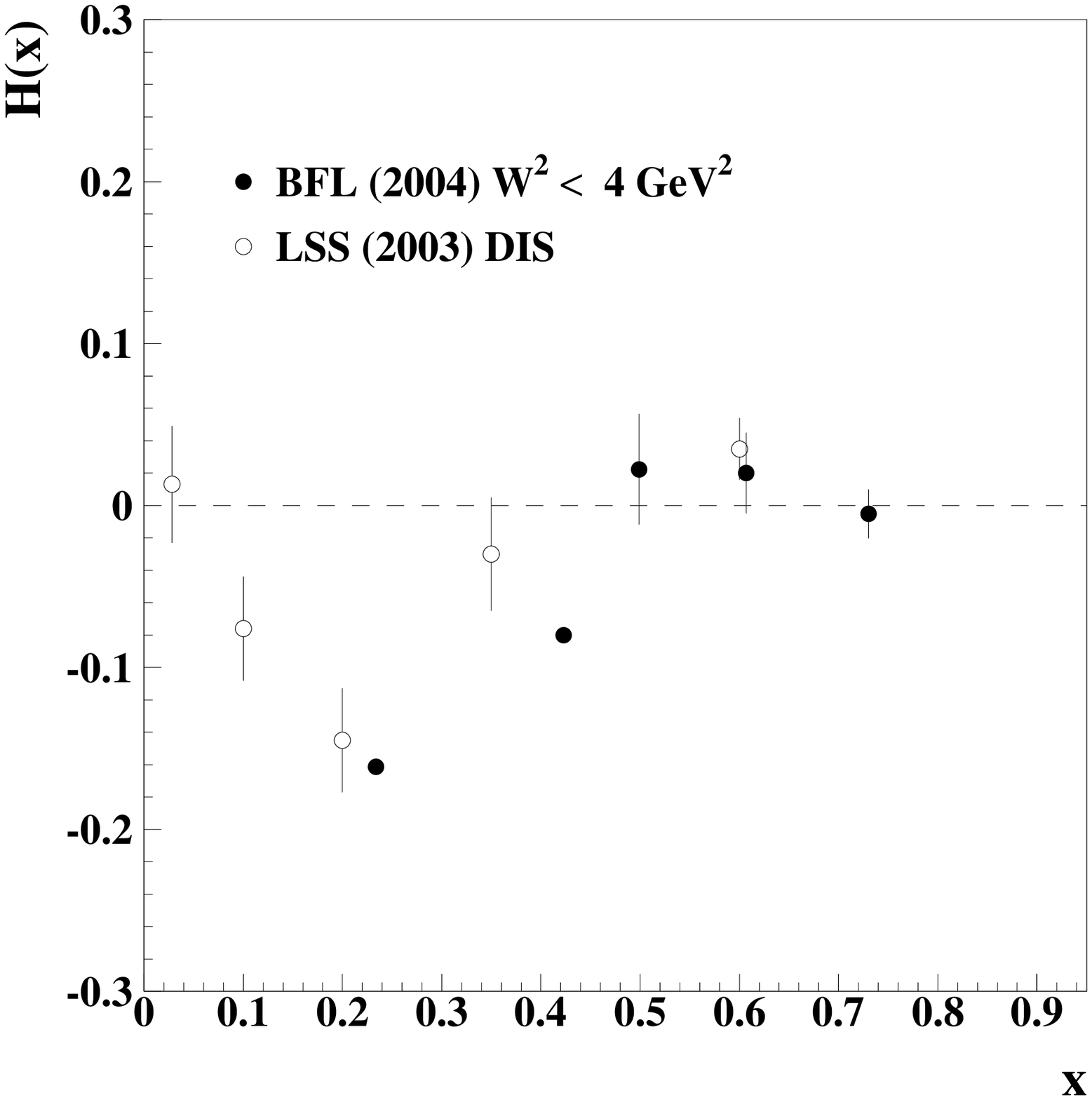}
\caption{HT coefficients for the unpolarized (left) and polarized (right) case.
         The full points represent the values in the resonance region, while 
         the empty points and the curves are related to the DIS region.}
\label{HTcontr}
\end{figure}

For unpolarized scattering it has been found that TMC and LxR diminish
considerably the space left for HT contributions.
The contribution of TMC is large at the largest values of $Q^2$ because these
correspond also to large $x$ values.
Moreover, the effect of TMC is larger than the one of LxR.
Similarly, in polarized scattering the inclusion of TMC and LxR decreases
the ratio $R_{pol}^{LT}$.
However, in this case these effects are included almost completely
within the error bars.
Clearly, duality is strongly violated at $Q^2 < 1.7$ GeV$^2$.
The present mismatch between the unpolarized and polarized low $Q^2$
behavior might indicate that factorization is broken differently for the two
processes, and that the universality of quark descriptions no longer holds.
The discrepancy from unity of the ratios already presented is interpreted in
terms of HTs.
In Figs.~\ref{theorx},\ref{HTcontr} the question of the size of the HT 
corrections is addressed explicitely.
For $F_2$, they are defined as:
\begin{equation}
\label{CHT}
H(x,Q^2) = Q^2 \left(F_2^{\mathrm{res}}(x,Q^2) - F_2^{\mathrm{LT}} \right)
\;\;
;
\;\;
C_{HT}(x) = \frac{H(x,Q^2)}{F_2^{pQCD}(x/Q^2)}
\equiv Q^2 \frac{F_2^{\mathrm{res}}(x,Q^2) - F_2^{LT}}{F_2^{\mathrm{LT}}}
\label{CHT1}
\end{equation}
A similar expression is assumed for $g_1$.
$C_{HT}$ is the so-called factorized form obtained by assuming that
the $Q^2$ dependences of the LT and of the HT parts are similar and therefore
they cancel out in the ratio. Although the anomalous
dimensions of the HT part could in principle be different, such a discrepancy
has not been found so far in accurate analyses of DIS data.
The HT coefficient, $C_{HT}$ has been evaluated for the three cases listed
also in Fig.\ref{theor1}, namely with respect to the NLO pQCD calculation, to
NLO+TMC and to NLO+TMC+LxR.
The values of $1 + C_{HT}/Q^2$ are plotted in Fig.\ref{theorx} (left panel)
as a function of the average value of $x$ for each spectrum. 
One can see that the NLO+TMC+LxR analysis yields very small values
for $C_{HT}$ in the whole range of $x$.
Furthermore, the extracted values are consistent with the ones obtained in 
Ref.\cite{SIMO1} using a different method, however the
present extraction method gives more accurate results.

In the polarized case (right panel) the HTs are small within the given
precision, for $Q^2 > 1.7$ GeV$^2$, but they appear to drop dramatically below
zero for lower $Q^2$ values.
The inclusion of TMC and LxR renders these terms consistent with zero at the
larger $Q^2$ values, but it does not modify substantially their behavior at
lower $Q^2$.
From a comparison with results of ratio including phenomenological
parametrizations~\cite{BFL} that includes some extra non perturbative 
behaviors it's possible to see that their effect seems not be large.

As mentioned in \cite{Ale2}, an accurate extraction of the $Q^2$ dependence 
is fundamental.
The results shown in \cite{BFL} have been extendend with the recent results
available in literature for the unpolarized and polarized case.
In Fig.~\ref{HTcontr}, the higher twist coefficients of the present extraction
in the resonance region are compared with all existing results of HT 
coefficients calculated in the DIS region.
For the unpolarized case there is the BCDMS evaluation \cite{BCDMS} (already 
shown in \cite{BFL}) and the new MRST calculation \cite{MRST}.
The HT coefficients have been calculated following the factorization formula, 
displayed in Eq.~\ref{CHT}, which can be expressed as
$F_2^{LT+HT} = F_2^{LT} \cdot (1 + C(x)/Q^2)$.
For the polarized one the only data available \cite{LSS} are using the 
additive formula, for which
$F_2^{LT+HT} = F_2^{LT} + H(x)/Q^2$
In the expression of $C(x)$ and $H(x)$ there is no $Q^2$ dependence hidden.
A different behavior for the unpolarized and polarized HT terms is evident. 
In details, for the unpolarized case in the region of high 
$x$ there is a big discrepancy between the HT terms in the resonance region 
($C_{res}(x)$) and in the DIS region ($C_{DIS}(x)$). 
In the polarized case, at high $x$ this comparison is little
bit complicated, due to the fact that there is only one point with $x>$0.6 
in the resonance region ($H_{res}(x)$) and no value for the DIS region 
($H_{DIS}(x)$).   

\bibliographystyle{aipproc}   

\bibliography{sample}

\begin{thebibliography}{}
\bibitem{BG} E.D.~Bloom and F.J.~Gilman, 
             \Journal{\PRL}{25}{1140}{1970};
             \Journal{\PRD}{4}{2901}{1971}.
\bibitem{CEBAF} I.~Niculescu {\em et al.}, 
                \Journal{\PRL}{85}{1186}{2000}.
\bibitem{whit} L.W.~Whitlow {\em et al.}, \Journal{\PLB}{282}{475}{1992}.
\bibitem{E143} E143 Coll., K.~Abe {\em et al.}, 
               \Journal{\PRD}{58}{112003}{1998}.
\bibitem{HERMESPRL} HERMES Coll., A.~Airapetian {\em et al.}, 
               \Journal{\PRL}{90}{092002}{2003}.
\bibitem{Ale} A.~Fantoni, \Journal{\EPJ}{A17}{385}{2003}.
\bibitem{BFL} N.~Bianchi, A.~Fantoni \& S.~Liuti,
              \Journal{\PRD}{69}{014505}{2004}.
\bibitem{Rob} R.~G.~Roberts, \Journal{\EPJ}{C10}{697}{1999}.
\bibitem{SIMO1} S.~Liuti, R.~Ent, C.E.~Keppel \& I.~Niculescu, 
                \Journal{\PRL}{89}{162001}{2002}.
\bibitem{Ale2}  A.~Fantoni, \Journal{\EPJ}{A}{}{2005} to be published.
\bibitem{MRST} A.D.~Martin {\em et al.}, hep-ph/0407311.
\bibitem{BCDMS} M.~Virchaux \& A.~Milsztajn, \Journal{\PLB}{274}{221}{1992}.
\bibitem{LSS} E.~Leader {\em et al.}, \Journal{\PRD}{67}{074017}{2003}.
\bibitem{Alekhin} S.I.~Alekhin, \Journal{\PRD}{68}{014002}{2003};
                  arXiv:hep-ph/0212370.

\end{thebibliography}

\IfFileExists{\jobname.bbl}{}
 {\typeout{}
  \typeout{******************************************}
  \typeout{** Please run "bibtex \jobname" to optain}
  \typeout{** the bibliography and then re-run LaTeX}
  \typeout{** twice to fix the references!}
  \typeout{******************************************}
  \typeout{}
 }

\end{document}